# Record-high-$T_c$ elemental superconductivity in scandium

Jianjun Ying,[1,*] Shiqiu Liu,[1] Qing Lu,[2] Xikai Wen,[1] Zhigang Gui,[1] Yuqing Zhang,[1] Xiaomeng Wang,[2] Jian Sun[2,†] and Xianhui Chen[1,3,4,‡]

[1]Department of Physics, and CAS Key Laboratory of Strongly Coupled Quantum Matter Physics, University of Science and Technology of China, Hefei, Anhui 230026, China

[2] National Laboratory of Solid State Microstructures, School of Physics and Collaborative Innovation Center of Advanced Microstructures, Nanjing University, Nanjing, 210093, China

[3]CAS Center for Excellence in Quantum Information and Quantum Physics, Hefei, Anhui 230026, China.

[4]Collaborative Innovation Center of Advanced Microstructures, Nanjing University, Nanjing 210093, China.

*E-mail: yingjj@ustc.edu.cn
†E-mail: jiansun@nju.edu.cn
‡E-mail: chenxh@ustc.edu.cn

**Elemental materials provide clean and fundamental platforms for studying superconductivity. However, the highest superconducting critical temperature ($T_c$) yet observed in elements has not exceeded 30 K. Discovering elemental superconductors with a higher $T_c$ is one of the most fundamental and challenging tasks in condensed matter physics. In this study, by applying high pressure up to approximately 260 GPa, we demonstrate that the superconducting transition temperature of elemental scandium (Sc) can be increased to 36 K, which is a record-high $T_c$ for superconducting elements. The pressure dependence of $T_c$ implies the occurrence of multiple phase transitions in Sc, which is in agreement with previous X-ray diffraction results. Optimization of $T_c$ is achieved in the Sc-V phase, which can be attributed to the strong coupling between $d$ electrons and moderate-frequency phonons, as suggested by our first-principles calculations. This study provides insights for exploring new high-$T_c$ elemental metals.**

Elemental solids are the simplest material systems and provide fundamental platforms for the study of various physical properties. Superconductivity is an intriguing phenomenon that has been observed in over 50 elements at ambient and high pressure. Although the superconducting critical temperatures ($T_c$) of most elements are relatively low, several elements exhibit $T_c$ values near or above 20 K [1,2], including sulfur ($T_c$ of ~17 K at 220 GPa) [3], vanadium ($T_c$ of ~17 K at 120 GPa) [4], lithium ($T_c$ of ~15–20 K at 30 GPa) [5,6], scandium (Sc) and yttrium (Y) ($T_c$ of ~20 K at 100 GPa) [7,8], calcium (Ca) ($T_c$ of ~21–29 K at 220 GPa) [9], and titanium (Ti) ($T_c$ of ~26 K at 240

GPa) [10,11]. Although metallic hydrogen has been proposed to be a room-temperature superconductor [12,13], its synthesis remains a major challenge and is under debate [14–16]. Despite the significant increase in $T_c$ for some elements at high pressure, the highest recorded $T_c$ in elemental solids is still below 30 K to date, and searching for higher $T_c$ in elements remains an important challenge.

As the first 3$d$-transition element, Sc is often grouped with rare-earth metals due to its similar chemical properties. It forms a hexagonal close-packed structure (Sc-I) at ambient pressure and adopts an incommensurate host–guest structure (Sc-II) above 23 GPa [17]. Interestingly, superconductivity appears and rapidly increases with increasing pressure in Sc-II. A maximum $T_c$ of nearly 19.6 K is reached at approximately 100 GPa; however, $T_c$ suddenly drops when the structure changes to Sc-III above 104 GPa [8]. The relatively high $T_c$ is attributed to $s$-$d$ transfer, similar to that of other rare earth and transition elements [8]. With further increasing pressure, Sc-III turns into Sc-IV above approximately 140 GPa and into Sc-V above approximately 240 GPa [18]. However, the structures of Sc-III and Sc-IV are still not fully resolved. Recent theoretical work suggests that the Ccca-20 phase is a likely candidate for the observed Sc-III phase, whereas the structure of Sc-IV may consist of random stacking of different structural units [19]. The structure of Sc-V has been demonstrated by X-ray diffraction (XRD) experiments to consist of 6-screw helical chains with a hexagonal lattice (space group $P6_122$) [18], which is unique and has not been observed in other elements. Despite its rich high-pressure phase diagram, the physical properties of the high-pressure phases have not yet been well explored, especially the superconducting properties of the Sc-IV and Sc-V phases.

In this letter, we found that $T_c$ can be increased from 18 K at 130 GPa to 28 K at 220 GPa for the Sc-IV phase. When the structure becomes Sc-V, $T_c$ can be increased to 36 K, which is the highest recorded $T_c$ in superconducting elements. Our first-principles calculations indicate that Sc-V is a typical phonon-mediated superconductor, and the strong coupling between $d$ electrons and moderate-frequency phonons is responsible for the high $T_c$.

Samples of 99.9% purity Sc were loaded into a diamond anvil cell. To prevent the reaction of Sc with air and water, all preparation processes were performed in an argon glovebox. To achieve pressures above 200 GPa, diamond anvils with various culets (30 or 50 µm) were used for high-pressure transport measurements. Four triangular 200 nm gold (with a 10 nm chromium underlayer) were deposited on the surface of the culet of the diamond anvil as the inner electrode. The electrodes were insulated from the rhenium gasket with a $c$-BN/epoxy mixture. A 15–20 µm diameter hole was drilled in the center of the gasket as the sample chamber. A small piece of Sc was cut from the ingots and then densely compressed into the sample chamber without any pressure medium. Miniature diamond anvil cells [20] or diamond anvil cells manufactured by the HMD corporation were used for high-pressure experiments. The pressure was calibrated using the shift of the diamond anvil Raman at room temperature, and the pressure difference was approximately 10–15 GPa above 200 GPa for different locations in the sample. Transport measurements were performed using the Physical Property Measurement System (Quantum Design) or TeslatronPT system (Oxford Instruments) with a 100 µA current applied to the sample. Details of our theoretical methods are described in the supplemental material [21].

To investigate the superconductivity of elemental Sc at high pressure, we performed high-pressure resistance measurements. The temperature dependence of the resistance for Sc under various pressures is presented in Fig. 1. Superconductivity appears in Sc-II, and $T_c$ gradually increases to 19 K with increasing the pressure to approximately 100 GPa, which is consistent with previous results [8]. $T_c$ slightly decreases by further increasing the pressure, which may be due to the structural phase transition from Sc-II to Sc-III. At pressures above 130 GPa, $T_c$ can be increased from 18 K at 130 GPa to 28 K at approximately 220 GPa in Sc-IV. With further increasing the pressure, superconductivity with a $T_c$ up to 36 K emerges when the structure possibly changes to Sc-V, as illustrated in Fig. 1(e) and Fig. S3 in the supplemental material [21]. This is the highest observed $T_c$ of any element. The superconducting phase diagram of Sc at high pressure is mapped in Fig. 2. The pressure dependence of $T_c$ suggests multiple phase transitions and the phase boundary determined by $T_c$ in our experiments is consistent with previous XRD results [18].

Compared with the previous high-pressure ac susceptibility work on Sc [8], the critical pressure is nearly 20 GPa lower in the Sc-II phase, which may be due to the pressure increase at low temperatures in our high-pressure resistance measurements. The $T_c$ determined by ac susceptibility is much lower than our results in the Sc-III phase, and the transition in the ac susceptibility is rather broad, as illustrated in Fig. S4 in the supplemental material [21]. This may be due to the filamentary/fractional superconductivity in the Sc-III phase, which occurs when the resistance starts to drop; however, bulk superconductivity can only be achieved at much lower temperatures.

In the Sc-V phase, $T_c$ reaches its maximum value of 36 K and is nearly unchanged with further compression. Our high-pressure superconducting phase diagram for Sc indicates that $T_c$ is highly correlated with the crystal structure. The neighboring elements of Sc in the periodic table (Ca, Ti, and Y) also exhibit relatively high $T_c$ at high pressure, suggesting the critical role of *s-d* charge transfer in these compressed elements. The exceptionally high $T_c$ observed in the Sc-V phase indicates that the crystal structure must also be taken into consideration for pursuing high $T_c$ in elemental superconductors. Because we only performed high-pressure resistance measurements, it is unclear whether the superconductivity is filamentary/fractional superconductivity for the Sc-IV and Sc-V phases. High-pressure magnetic and/or heat capacity measurements should be performed in future work to examine the bulk superconductivity, although these techniques are challenging above 200 GPa.

Figure 3 illustrates the superconducting transition under different magnetic fields for Sc at various pressures. We can estimate the upper critical field ($H_{c2}$) from these resistivity curves. $H_{c2}$ exhibits a linear dependence on $T_c$, as illustrated in Fig. S5(a) in the supplemental material [21]. In the weak-coupling Bardeen–Cooper–Schrieffer theory, the upper critical field at $T = 0$ K can be determined by the Werthamer–Helfand–Hohenberg equation [34] $\mu_0 H_{c2}(0) = 0.693[-(d\mu_0 H_{c2}/dT)]_{T_c} T_c$. We can deduce $\mu_0 H_{c2}(0)$ for Sc at various pressures, as displayed in Fig. S5(b) in the supplemental material [21]. $\mu_0 H_{c2}(0)$ for the Sc-IV phase is approximately 40 T and decreases to approximately 30 T for the Sc-V phase. The Helfand–Werthamer (HW) theory is often applied to anisotropic superconductors, especially to two-band clean materials [34,35]. In the HW scheme, the sudden change in the $[-(d\mu_0 H_{c2}/dT)]_{T_c}/T_c$ value can be attributed to the change in the Fermi surface topology [36]. The pressure dependence of $[-(d\mu_0 H_{c2}/dT)]_{T_c}/T_c$ is presented in Fig. S5(b) in the supplemental material [21]. We observe that $[-(d\mu_0 H_{c2}/dT)]_{T_c}/T_c$ exhibits anomalies at approximately 130 and 230 GPa, as demonstrated in Fig. S5(b), which correspond to the Sc-III to Sc-IV and Sc-IV to Sc-V structural phase transitions, respectively.

To obtain insights into the mechanism of the high $T_c$ in the Sc-V phase, we performed first-principles calculations on the electronic structures and superconducting properties of Sc-V at 230 GPa, as illustrated in Figs. 4 and 5. Fig. 4(a) illustrates the calculated Brillouin zone and selected high-symmetry-point path for the electronic band and phonon dispersion calculations. Fig. 4(b) indicates that four bands cross the Fermi level, most of which are composed of electrons from the unfilled $d$ orbitals. These $d$ electrons contribute significantly to the total density of states at the Fermi level. Particularly, from the band dispersions in Fig. 4(b), we can see van Hove singularities on the bands on the Fermi level near the K and H points as well as on the K→Γ line, which may further increase the total density of states. Fig. 4(c) indicates that there are multiple cylindrical Fermi pockets along the Γ→A path and several ellipsoidal Fermi pockets around the H, K, M, L and 1/2AL points. The variety of the Fermi surface indicates complex nesting, which further suggests the possibility of superconductivity.

Therefore, we performed electron–phonon coupling calculations for the Sc-V phase. Fig. 5(a) illustrates the calculated $\frac{\gamma_{qj}}{\omega_{qj}}$ resolved phonon spectrum, phonon density of states (PHDOS), Eliashberg function $\alpha^2 F(\omega)$, and accumulated electron–phonon coupling strength $\lambda(\omega)$. The absence of imaginary frequencies demonstrates the dynamic stability of the Sc-V phase at 230 GPa. We can classify the phonon spectrum into three zones: a low-frequency zone below 100 cm$^{-1}$, a moderate-frequency zone from 100 to 400 cm$^{-1}$, and a high-frequency zone above 400 cm$^{-1}$. The strong coupling mainly originates from the moderate-frequency zone, which contributes approximately 94.5% of the electron–phonon coupling strength λ, especially around the M, K, L, and H points (also see Fig. S7 in the supplemental material [21]). The small value of PHDOS in the low-frequency zone and the inverse relationship between λ and ω suppress the coupling from low-frequency phonons and high-frequency phonons, respectively.

The final results of our calculations for Sc at 230 GPa are λ = 1.27 and $T_c$ = 33.0 K. The high $T_c$ can be attributed to the strong coupling between electrons from unfilled *d* orbitals and phonons from the moderate-frequency zone. We also calculated λ and $T_c$ at 250 and 270 GPa, as illustrated in Fig. 5(b). From 230 to 270 GPa, λ and $T_c$ of Sc-V do not change significantly, remaining at 1.27–1.28 and 33.0–34.9 K, respectively. The calculated value of λ is close to that determined by analysis of the normal-state resistance, as presented in the supplemental material [21]. The variation of $T_c$ is very small and consistent with the experimental results. These results indicate that Sc-V is a typical phonon-mediated superconductor, which is different from δ-Ti, where the correlation effect of *d* electrons significantly influences $T_c$ [11].

In conclusion, we performed high-pressure transport measurements on elemental Sc up to approximately 260 GPa using diamond anvil cells and observed a record-high superconducting $T_c$ of 36 K in the Sc-V phase. The pressure dependence of $T_c$ suggests multiple phase transitions, which is consistent with previous XRD results. The results of our first-principles calculations are in good agreement with the experimental results, indicating that Sc-V is a conventional phonon-mediated superconductor. The strong coupling between *d* electrons and moderate-frequency phonons is responsible for the high $T_c$. Our findings provide an avenue for the design and synthesis of new high-$T_c$ superconductors among simple materials under extreme conditions.

 Acknowledgments

This work was supported by the National Natural Science Foundation of China (Grants Nos. 11888101, 11534010, 12125404, 11974162, and 11834006), CAS Project for Young Scientists in

Basic Research (Grant No. YBR-048), the National Key Research and Development Program of the Ministry of Science and Technology of China (Grant No. 2019YFA0704900), the Anhui Initiative in Quantum Information Technologies (Grant No. AHY160000), the Science Challenge Project of China (Grant No. TZ2016004), the Innovation Program for Quantum Science and Technology (Grant No. 2021ZD0302800), the Key Research Program of Frontier Sciences, CAS, China (Grant No. QYZDYSSWSLH021), the Strategic Priority Research Program of the Chinese Academy of Sciences (Grant No. XDB25000000), the Collaborative Innovation Program of Hefei Science Center, CAS, (Grant No. 2020HSC-CIP014) and the Fundamental Research Funds for the Central Universities (Grant Nos. WK3510000011 and 020414380191). The calculations were carried out using supercomputers at the High Performance Computing Center of Collaborative Innovation Center of Advanced Microstructures, the high-performance supercomputing center of Nanjing University.

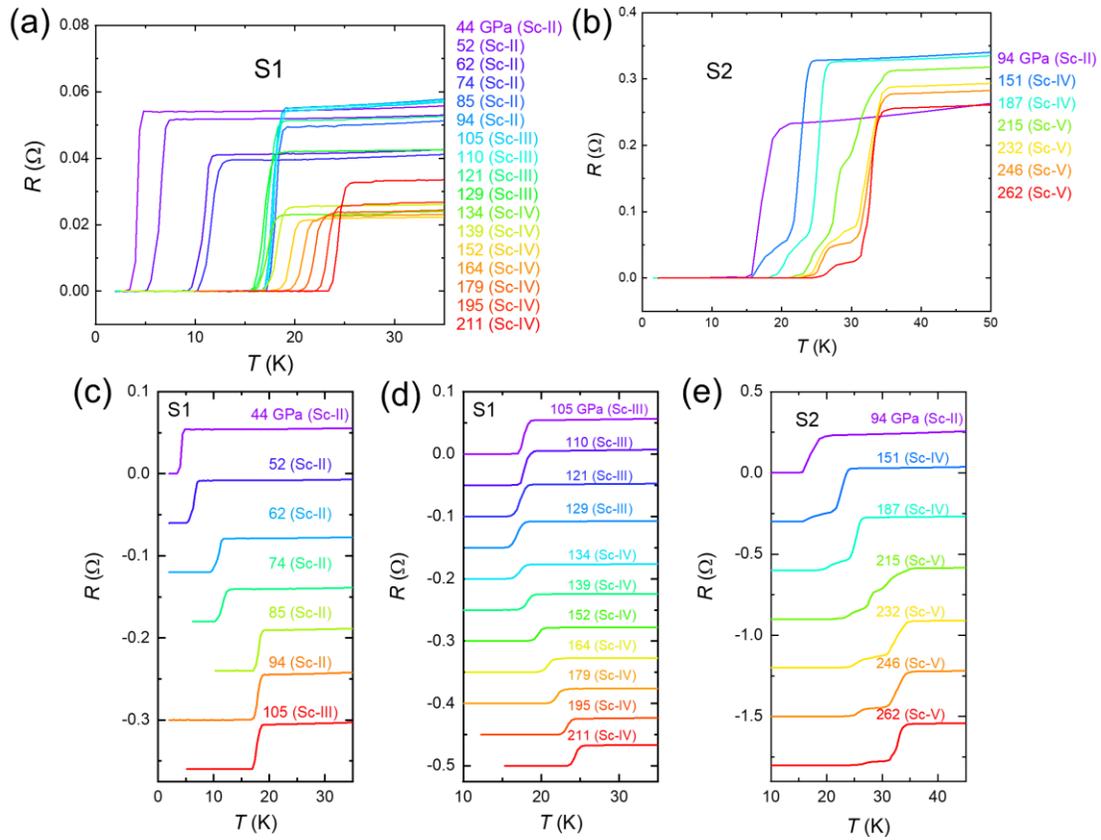

FIG. 1. (a), (b) Temperature dependence of resistance for samples S1 and S2. (c), (d), (e) Replot of resistance curves for S1 and S2. All the curves are shifted vertically for clarity. (c) $T_c$ gradually increases with increasing the pressure up to 100 GPa. (d) $T_c$ is slightly suppressed with increasing the pressure up to 130 GPa and then starts to increase again with further increasing the pressure. (e) Superconductivity up to 36 K emerges above approximately 220 GPa.

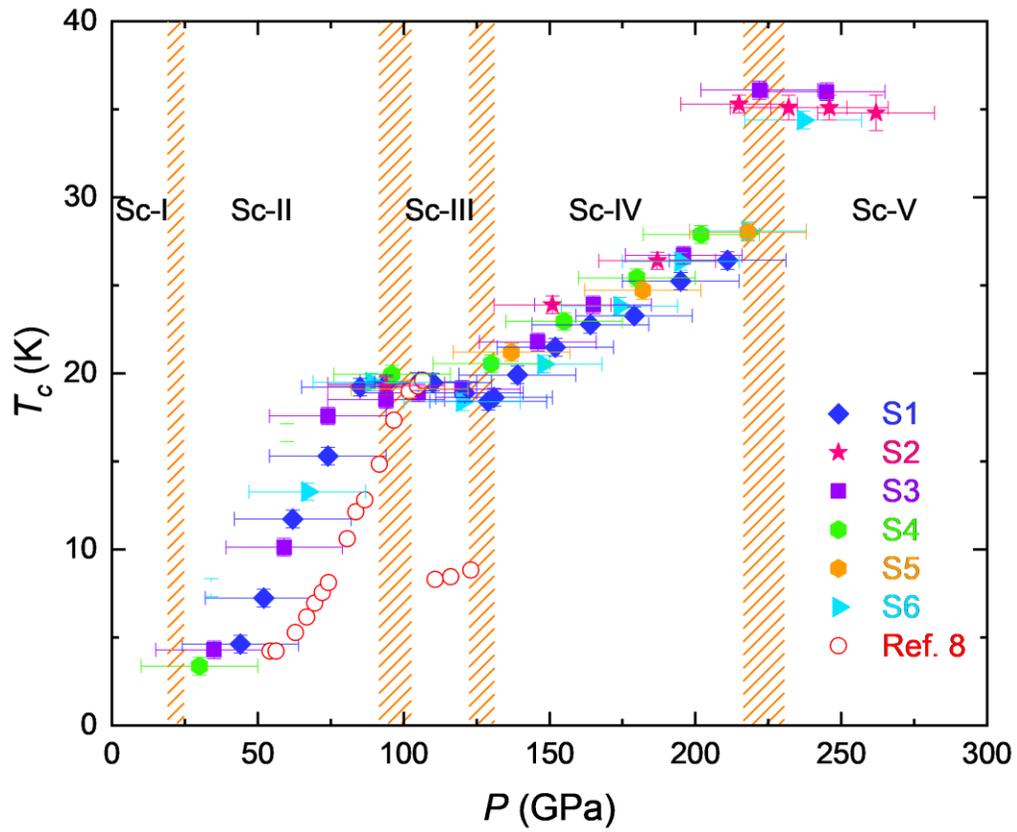

FIG. 2. Superconducting critical transition temperature $T_c$ of elemental scandium (Sc) at high pressure. The dashed areas represent the phase boundaries for compressed Sc with different crystal structures. $T_c$ is determined from the onset of the resistance transition.

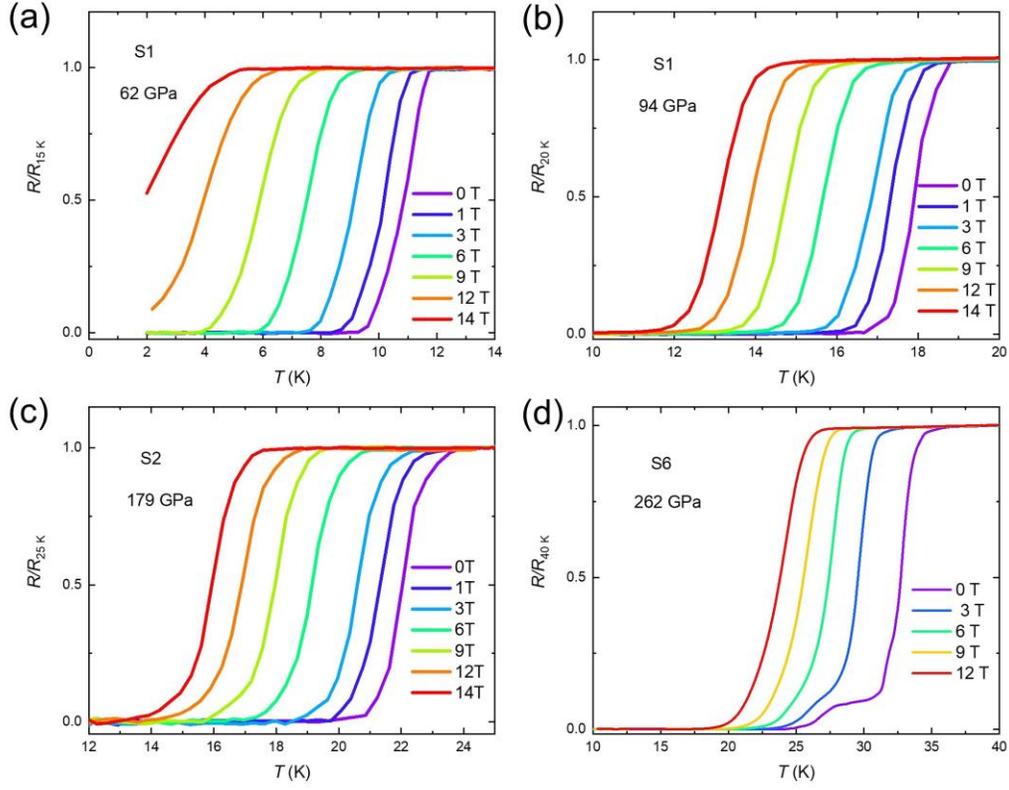

FIG. 3. Temperature dependence of the resistance of scandium (Sc) at various magnetic fields. The upper critical field $\mu_0 H_{c2}$ for various pressures can be extracted from these curves. $T_c$ is determined from the onset of the resistance transition, as displayed in Fig. S5(a).

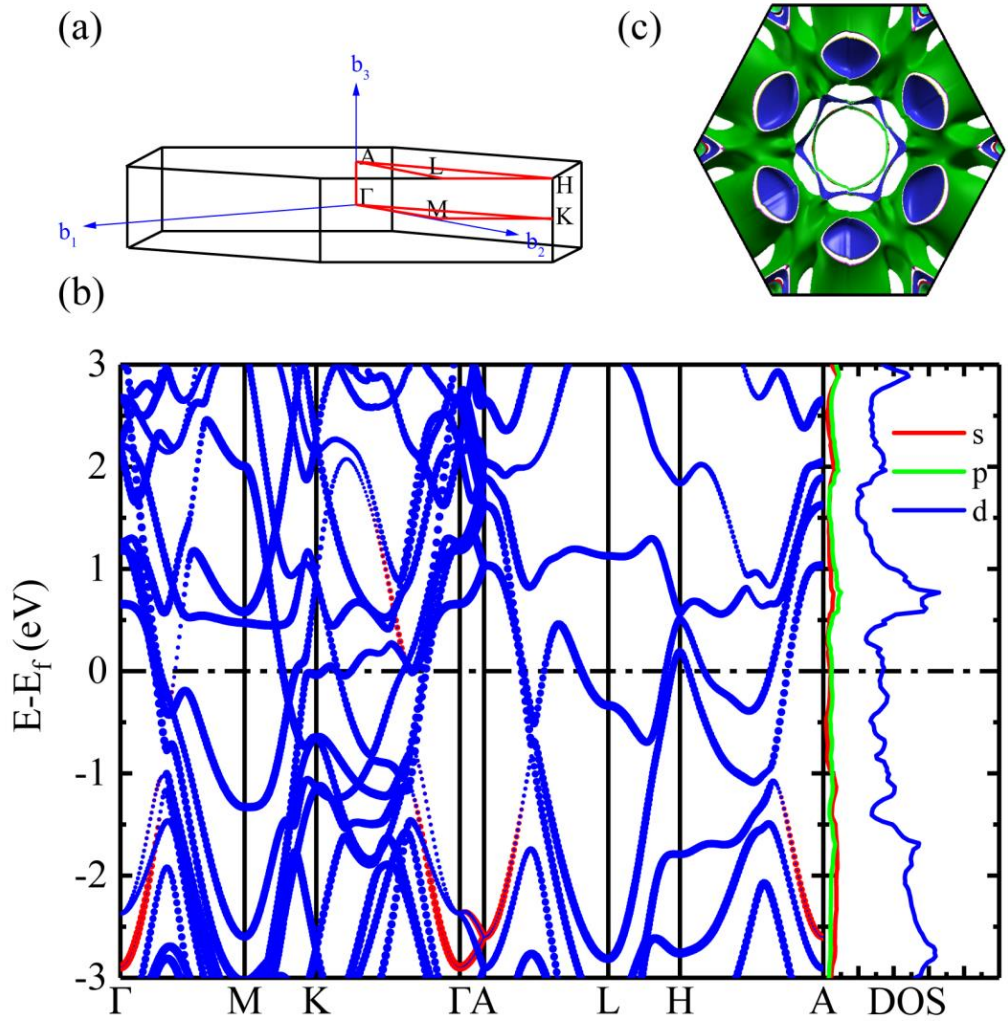

FIG. 4. Electronic properties of Sc-V at 230 GPa. (a) Reciprocal lattice. The Brillouin zone is surrounded by black lines, with red lines representing high-symmetry-point paths and the blue axis representing the reciprocal vector. (b) Projected band structure and density of states. Red, green, and blue dots/lines represent the contributions of electrons from *s*, *p*, and *d* orbitals, respectively. The size of the dots in the band dispersion is proportional to the contribution of electrons from related orbitals. The Fermi level is shifted to 0 eV, as labeled by the dash-dotted line. (c) Top view of the Fermi surface, which reveals hexagonal symmetry and complex nesting.

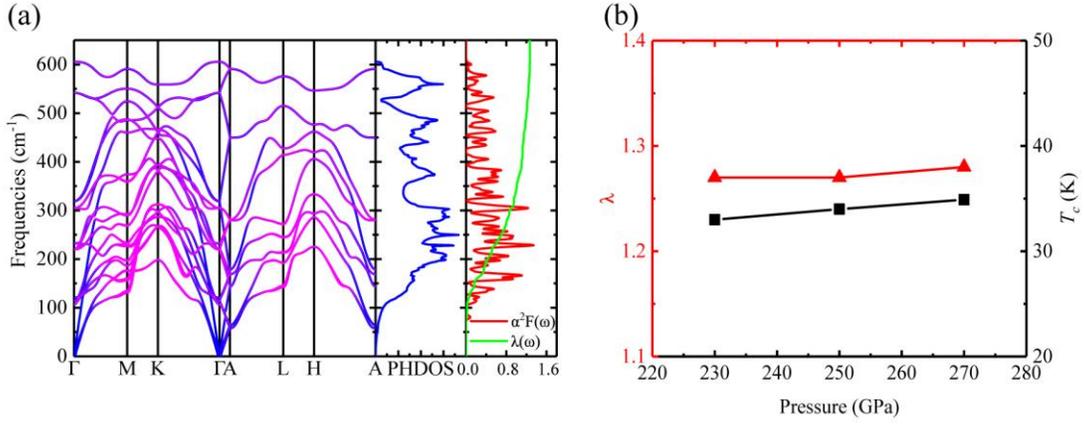

FIG. 5. Superconducting properties of Sc-V. (a) Calculated $\frac{\gamma_{qj}}{\omega_{qj}}$ resolved phonon spectrum, phonon density of states, Eliashberg function $\alpha^2 F(\omega)$, and accumulated electron–phonon coupling strength $\lambda(\omega)$ at 230 GPa. The size of the magenta circles is proportional to the value of the dimensionless parameter $\frac{\gamma_{qj}}{\omega_{qj}}$, which originates from $\alpha^2 F(\omega) = \frac{\hbar}{2\pi N(E_f)} \frac{1}{N_q} \sum_{qj} \frac{\hbar}{2\pi N(E_f)} \frac{1}{N_q} \frac{\gamma_{qj}}{\omega_{qj}} \delta(\omega - \omega_{qj})$ and evaluates the coupling in $\alpha^2 F(\omega)$. (b) Superconducting properties at 230, 250, and 270 GPa. The solid red triangles and solid black squares represent the electron–phonon coupling strength λ and transition temperature $T_c$, respectively.